\documentclass{PoS}
\newcommand{\charpm}[1]{\tilde\chi^\pm_{#1}}
\newcommand{\neut}[1]{\tilde\chi^0_{#1}}

\title{
Natural SUSY: LHC and Dark Matter direct detection experiments interplay}

\ShortTitle{Natural SUSY: LHC and Dark Matter direct detection experiments interplay}

\author{\speaker{D. Barducci}\\
        LAPTh, Universit\'e de Savoie Mont Blanc, CNRS, B.P.110, F-74941 Annecy-le-Vieux, France\\
        E-mail: \email{barducci@lapth.cnrs.fr}}

\author{A. Belyaev$^{a,b}$, A. Bharucha$^c$, W. Porod$^d$ and V. Sanz$^e$\\
       $^a$ School of Physics and Astronomy, University of Southampton, Highfield, Southampton SO17~1BJ, UK\\
       $^b$ Particle Physics Department, Rutherford Appleton Laboratory, Chilton, Didcot, Oxon OX11~0QX, UK\\
        $^c$ CNRS, Aix Marseille U., U. de Toulon, CPT, UMR 7332, F-13288, Marseille, France\\
        $^d$ Institut f\"ur Theoretische Physik und Astrophysik, Universit\"at  W\"urzburg, D-97074  W\"urzburg, Germany\\
       $^e$ Department of Physics and Astronomy, University of Sussex, Brighton BN1 9QH, UK
       }

\abstract{
 \begin{flushright}
  LAPTH-CONF-031/15\\
 \end{flushright}
 Natural SUSY scenarios with a low value of the $\mu$ parameter, are characterised by a higgsino-like dark matter candidate, and a compressed spectrum for the lightest higgsinos.
We explore the prospects for probing this scenario at the 13 TeV stage of the LHC via monojet searches, with various integrated luminosity options, and demonstrate how these results are affect by different assumptions on the achievable level of control on the experimental systematic uncertainties.
The complementarity between collider and direct detection experiments (present and future) is also highlighted.

}

\FullConference{18th International Conference From the Planck Scale to the Electroweak Scale \\
		25-29 May 2015\\
		Ioannina, Greece }

\begin{document}

\section{Introduction}

The lack of evidence for supersymmetric (SUSY) particles at the large hadron collider (LHC), and the measured value of the Higgs mass, raise the question of whether the remaining parameter space of the minimal supersymmetric standard model (MSSM) suffers or not from a high degree of fine tuning.

The fine tuning measure~\cite{Ellis:1986yg,Barbieri:1987fn}, defined as the sensitivity
of the electroweak (EW) scale to fractional variations in the fundamental parameters of a theory, can in principle be low even if SUSY scalar sparticles are rather heavy, for example in the hyperbolic branch of focus point regions of the minimal supergravity parameter space. Moreover, recent arguments~\cite{Baer:2013gva} pointed out that conventional fine tuning measures in SUSY scenarios can be grossly overestimated, by neglecting additional contribution arising from the ultra violet completion of the model. These terms can lead to large cancellations that can favour a low $\mu$ parameter, which will take as a definition of natural SUSY in this work.

In the case that $\mu \ll M_1,M_2$, where $M_{1,2}$ are the gauginos mass parameters, the lightest SUSY spectrum is characterised by three quasi degenerate states, $\neut{1,2}$ and $\charpm{1}$, which are nearly pure higgsinos.
The small mass splitting between them, makes this scenario challenging to be probed at the LHC, while the Higgsino nature of the lightest SUSY particle (LSP), favours a dark matter (DM) relic density which is below the Planck measurements~\cite{Adam:2015rua}, $\Omega h^2_{\rm Planck}$ = 0.1186$\pm$0.0020. 
This region of the MSSM parameter space can be quite precisely described, from the monojet perspective, in terms of just two parameters: the LSP mass, $m_{\neut{1}}$, and its mass splitting with respect to the next to LSP (NLSP), which is usually the lightest chargino, $\Delta M=m_{\charpm{1}}-m_{\neut{1}}$.

Is the purpose of this work, based on Ref.~\cite{Barducci:2015ffa}, to study the complementarity between collider and direct detection experiments in covering this \emph{natural} configuration of the MSSM.

\section{Natural SUSY spectrum and DM properties}

In the bases $(\tilde B^0,\tilde W^0, \tilde H^0_d,\tilde H^0_u)$ and $(\tilde W^0, \tilde H^0_d)$, where $\tilde B, \tilde W$ and $\tilde H$ are the SUSY partners of the standard model (SM) $B,W$ and H fields , the mass matrices of the neutralino and chargino sector of the MSSM are
\begin{equation}
M_{\neut{}}=
\left(
\begin{array}{c c c c}
M_1                      & 0                      & -M_Z s_\omega c_\beta & M_Z s_\omega s_\beta \\
0                        & M_2                    &  M_Z c_\omega c_\beta & -M_Z c_\omega s_\beta \\
-M_Z s_\omega c_\beta    & M_Z c_\omega c_\beta   &                       & -\mu                  \\
M_Z s_\omega s_\beta     &-M_Z c_\omega s_\beta   & -\mu                  & 0                  \\
\end{array}
\right)
\quad\,
M_{\charpm{}}=
\left(
\begin{array}{c c}
M_2                   & \sqrt{2} M_W s_\beta \\
\sqrt{2} M_W c_\beta  & \mu  \\
\end{array}
\right)
\end{equation}
where $M_1$ and $M_2$ are the soft SUSY breaking mass parameter for $\tilde B$ and $\tilde W$, $\mu$ is the Higgsino mass parameter, $c_\omega$ and $s_\omega$ are $\cos$ and $\sin$ of the Weinberg angle, $\tan\beta=v_2/v_1$ is the ratio of the vacuum expectation
values of two Higgs doublets, $s_\beta$, $c_\beta$ are $\sin \beta$ and $\cos \beta$ respectively and $m_Z$, $m_W$ are the masses of the SM gauge bosons $Z^0$ and $W^\pm$.

In the limit where $\mu \ll M_1,M_2$, the mass splitting between the LSP and the NLSP is given by:
\begin{eqnarray}
\Delta M & = & m_{\charpm{1}} -  m_{\neut{1}} \simeq \frac{M_Z^2}{2}\left(\frac{s_\omega^2}{M_1} +\frac{c_\omega^2}{M_2}\right)  + |\mu| \frac{\alpha(m_Z)}{\pi}\left(2+\ln\frac{m^2_Z}{\mu^2}\right),
\end{eqnarray}
where we have neglected corrections $\mathcal{O}(1/\tan{\beta})$ and have included the electromagnetic correction to the $\charpm{1}$ mass.
\begin{figure}[htb]
\includegraphics[width=0.48\textwidth]{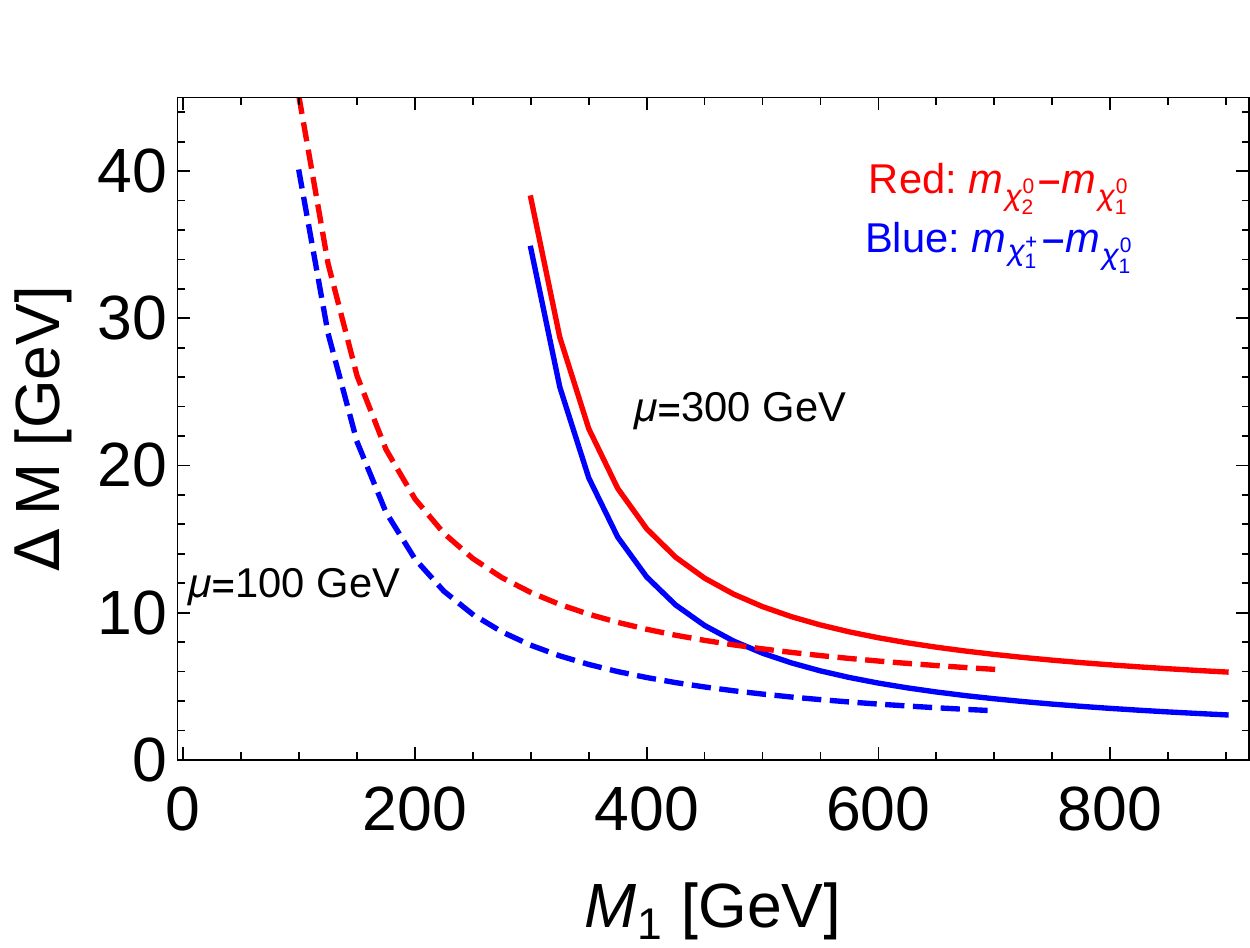}{}\hfill
\includegraphics[width=0.48\textwidth]{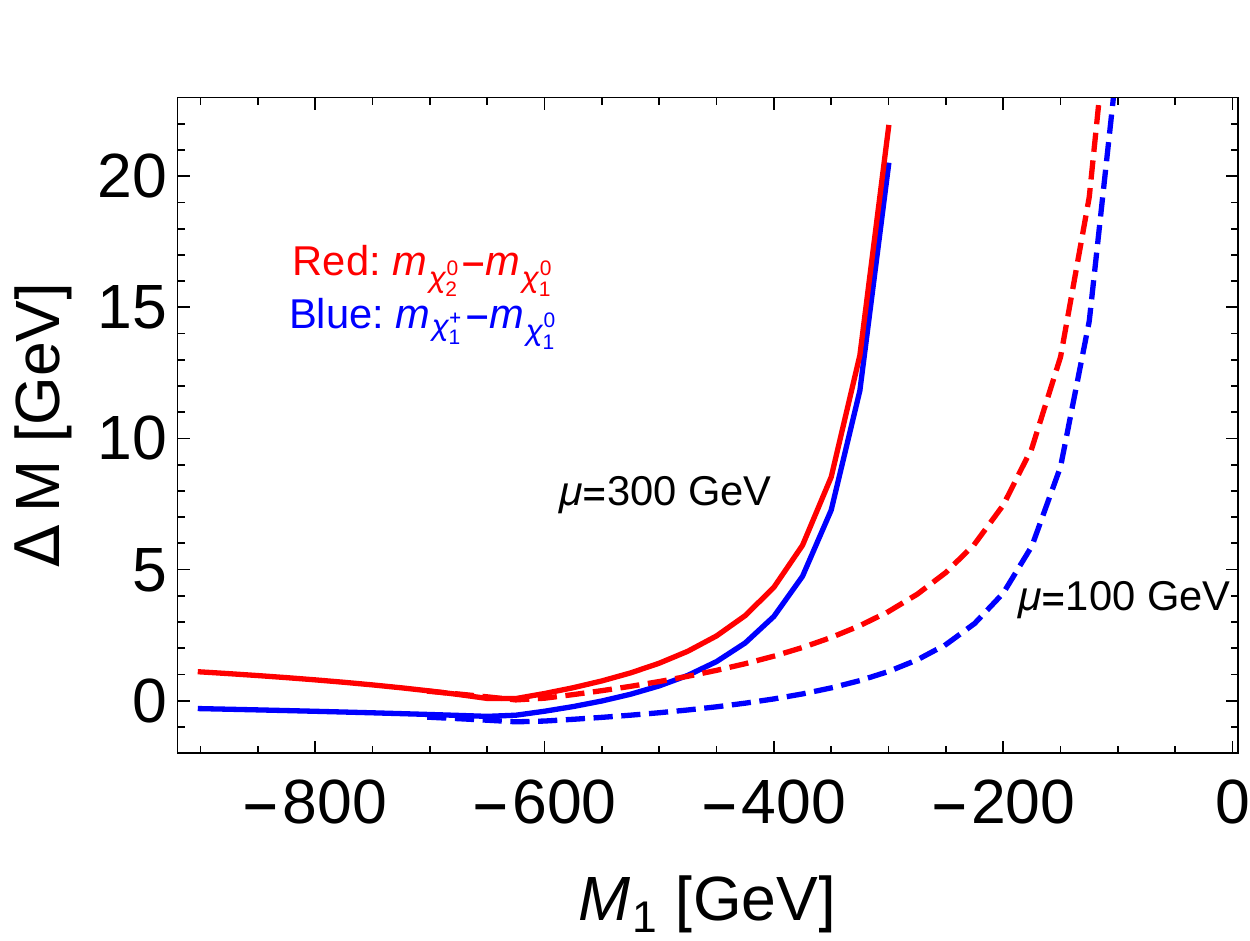}{}\\
\caption{$\charpm{1}-\neut{1}$ and $\neut{2}-\neut{1}$ mass splitting values as a  function of 
and $M_1$ for the case $M_1>0$ (left) and $M_1<0$ (right).}
\label{fig:m-dm_mu-m1}
\end{figure}

In order to efficiently work in the $m_{\neut{1}}$--$\Delta M$ parameter space, we have fixed $M_2$ to the value of 2 TeV, therefore decoupling $\neut{4}$ and $\charpm{2}$ which will be not considered anymore during our analysis, and explored the following range for $\mu$ and $M_1$
\begin{equation}
\mu=(100, 300) \textrm{ GeV}\qquad |M_1|-\mu=(0,600)~{\rm GeV}
\label{eq:param_space},
\end{equation} fixing $\tan\beta=5$. As shown in Fig.~\ref{fig:m-dm_mu-m1}, this choices provides a good control on the $\Delta M$ value,  that can be tuned by moving from higher to lower value the quantity $|M_1|-\mu$.

As mentioned, this choice of parameters makes the LSP a dominantly higgsino state, with a variable bino component, therefore causing the DM relic density, $\Omega h^2$, to be normally below the latest value measured by Planck, especially if $\mu\gg M_1$. 

\begin{figure}[!h]
\includegraphics[width=0.49\textwidth]{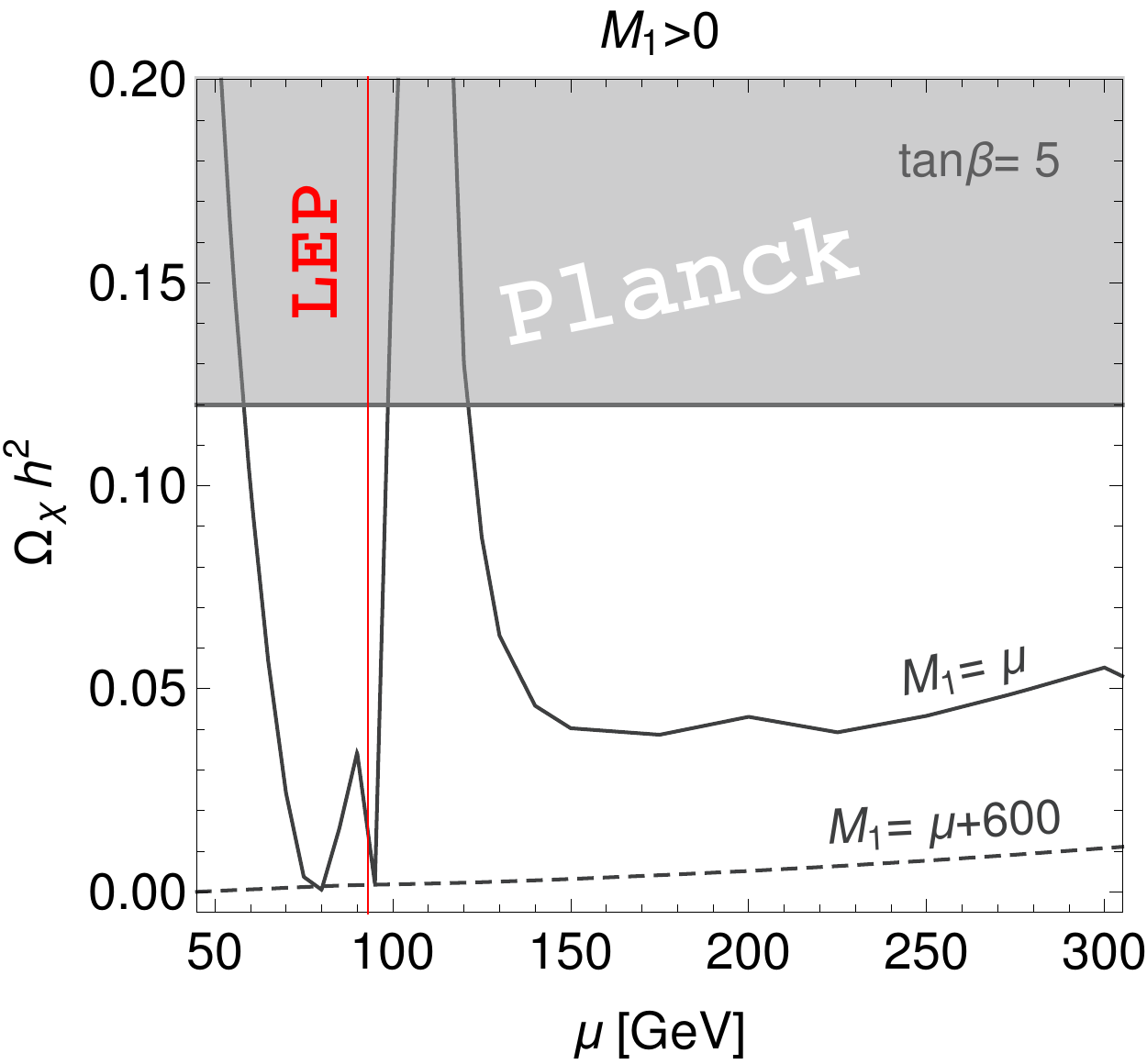}\hfill
\includegraphics[width=0.47\textwidth]{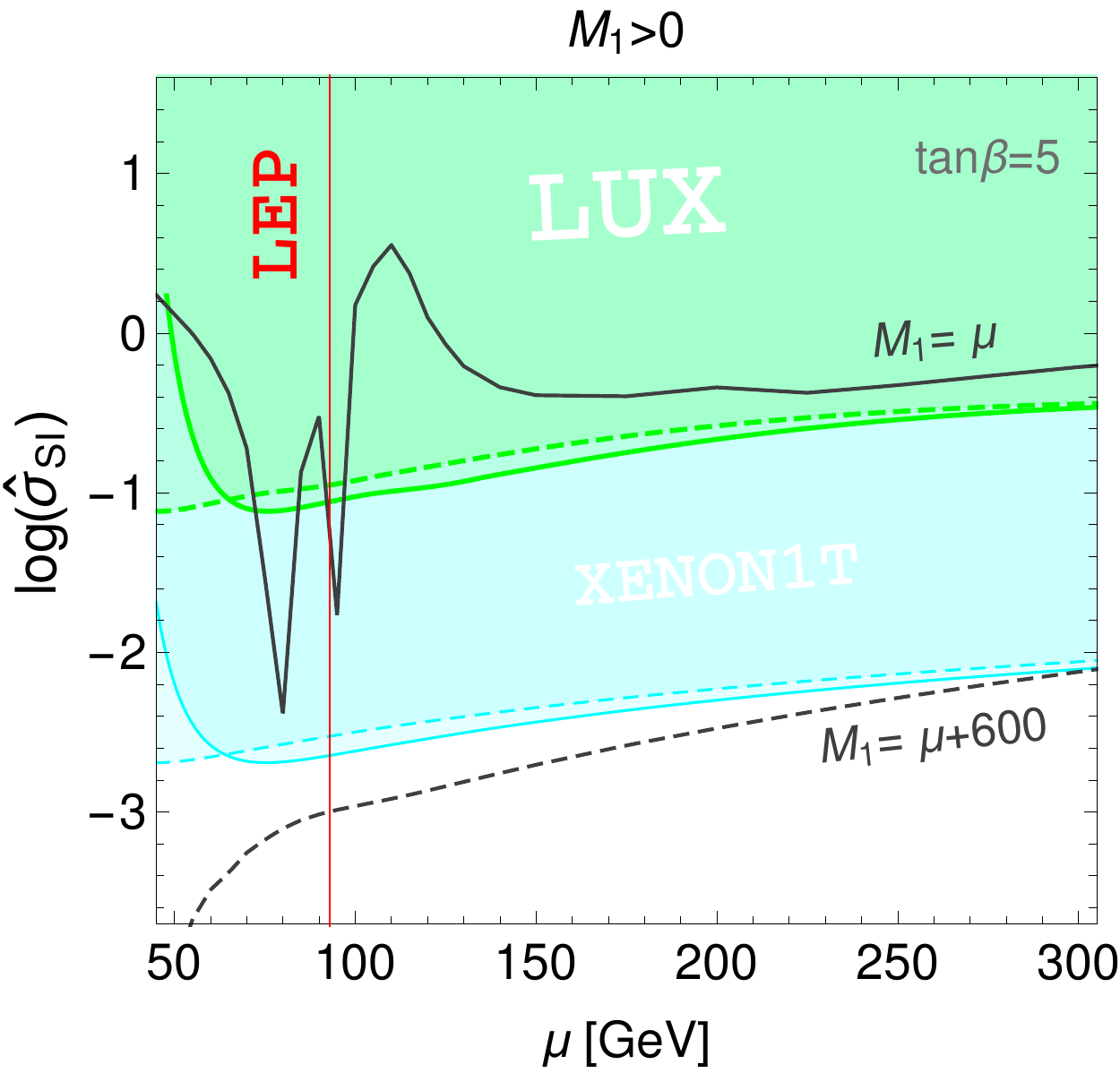}
\caption{Left panel: predicted value of the dark matter relic density $\Omega h^2$ as a function of $\mu$. The relic density measured by the Planck satellite, $\Omega h^2_{\rm Planck}$, is also shown for comparison, and the region excluded due to an overabundance of DM is indicated in gray.
Right panel: logarithm of the predicted value of the spin-independent annihilation cross section for DD $\hat{\sigma}_{\rm SI}=R_\Omega\,\sigma_{\rm SI}/(10^{-8}\,{\rm pb})$, 
 where $R_\Omega=\Omega/\Omega_{\rm Planck}$, as a function of $\mu$ (right). The excluded limit from LUX (green), as well as the projected exclusion from XENON1T (cyan) are also shown for comparison. Solid and dashed lines represent the exclusions for $M_1=\mu$ and $M_1=\mu$+600 GeV respectively. In both plots the LEP limit is inferred from the $\charpm{1}$ mass.
 } 
\label{fig:dm_rd-dd}
\end{figure}
This is illustrated in Fig.~\ref{fig:dm_rd-dd} (left panel), where $\Omega h^2$ is plot as a function of  $\mu$ for two values of $M_1-\mu$: 0 and 600 GeV, together with the region currently excluded due to an overabundance of DM by the Planck measurements.

For the same $M_1-\mu$ choices, we show in the right panel of Fig.~\ref{fig:dm_rd-dd} the predicted value of the spin-independent annihilation cross section for direct detection (DD), rescaled by the local relic density, $\hat{\sigma}_{\rm SI}=\sigma_{\rm SI}~\Omega/\Omega_{\rm Planck}$. As illustrated, for the choice $\mu=M_1$, the current results from the LUX experiment~\cite{Akerib:2013tjd} are already able to set a limit for low $M_1$, while future underground experiments, such as XENON1T~\cite{Aprile:2012zx}, will be able to probe a region with slightly higher $M_1$.

Interestingly then, DM DD experiments are able to probe regions of the natural SUSY parameter space with a high $\Delta M$, which is also the configuration easier to be tested at collider experiments, since the decay products of the NLSP can be hard enough to be detected. Conversely, in a lower $\Delta M$ regime, these decay products can become extremely soft, up to the point that they can become undetectable. In this situation then, one has usually to rely on a mono-object signature, {\it i.e.} a signature with a high $p_T$ object recoiling against missing transverse energy ($E_T^{\rm miss.}$).

\section{LHC phenomenology}

We now wish to investigate the complementarity between the LHC and DD experiments, anticipated in the previous Section.
We will focus on a monojet signature, {\it i.e.} the production of a pair of electroweakinos (EWinos) through the s-channel exchange of a SM EW gauge boson~\footnote{We assume all the squarks to be decoupled throughout the study.}, together with a hard QCD initial state radiation
\begin{equation}
p p \to \neut{a} \neut{b} j,\quad \neut{a,b}=\neut{1,2,3},\charpm{1}.
\label{eq:monojet}
\end{equation}
The main SM background for this signature consists of the irreducible $Zj,Z\to \nu\bar\nu$ process, while $Wj, W\to l\nu$ production gives a smaller contribution, when the lepton arising from the W decay is missed.

Both ATLAS~\cite{Aad:2015zva} and CMS~\cite{CMS:rwa} have performed studies of monojet signatures during the first run of the LHC, and we will here focus on the reinterpretation of the results at the 8 TeV LHC of the CMS search, moving then in presenting projection at the 13 TeV stage of the CERN machine for a possible upgrade of this analysis.

\subsection{8 TeV reinterpretation}

We have generated signal samples for the process given in eq.~(\ref{eq:monojet}) through \verb#MadGraph v.1.5.11#~\cite{Alwall:2014hca}. Parton shower, hadronization and decay of unstable particles have been performed via \verb#PYTHIA v6.4#~\cite{Sjostrand:2006za}, while \verb#Delphes v.3.2.0#~\cite{deFavereau:2013fsa} has been employed for a fast detector simulation, together with the \verb#Fastjet#~\cite{Cacciari:2011ma} package, for jet reconstruction with an anti-$k_T$~\cite{Cacciari:2008gp} algorithm.

We have then applied the following signal region selections from the CMS monojet analysis to our signal samples  
\begin{itemize}
\item[-] Leading jet with $p_T>$~110 GeV and
$|\eta|<$ 2.4.
\item[-] Events with more than two jets with $p_T>$ 30 GeV and $|\eta|<$ 4.5 are discarded together with events
with $\Delta\phi(j_1,j_2)<$~2.5, where $j_1$ and $j_2$ are the leading and sub-leading jets, to reduce QCD background.
\item[-] Veto on events with electrons or muons
with $p_T>$ 10 GeV and events with tau jets with $p_T>$~20 GeV and $|\eta|<$ 2.3, to suppress $W$ production background
\item[-]Finally the analysis was performed in 7 regions with an increasing requirement of $E_T^{\rm miss.}$: $E_T^{\rm miss.}>$ 250, 300, 350, 
400, 450, 500 and 550 GeV.
\end{itemize}

\begin{figure}[h!]
\includegraphics[width=0.46\textwidth]{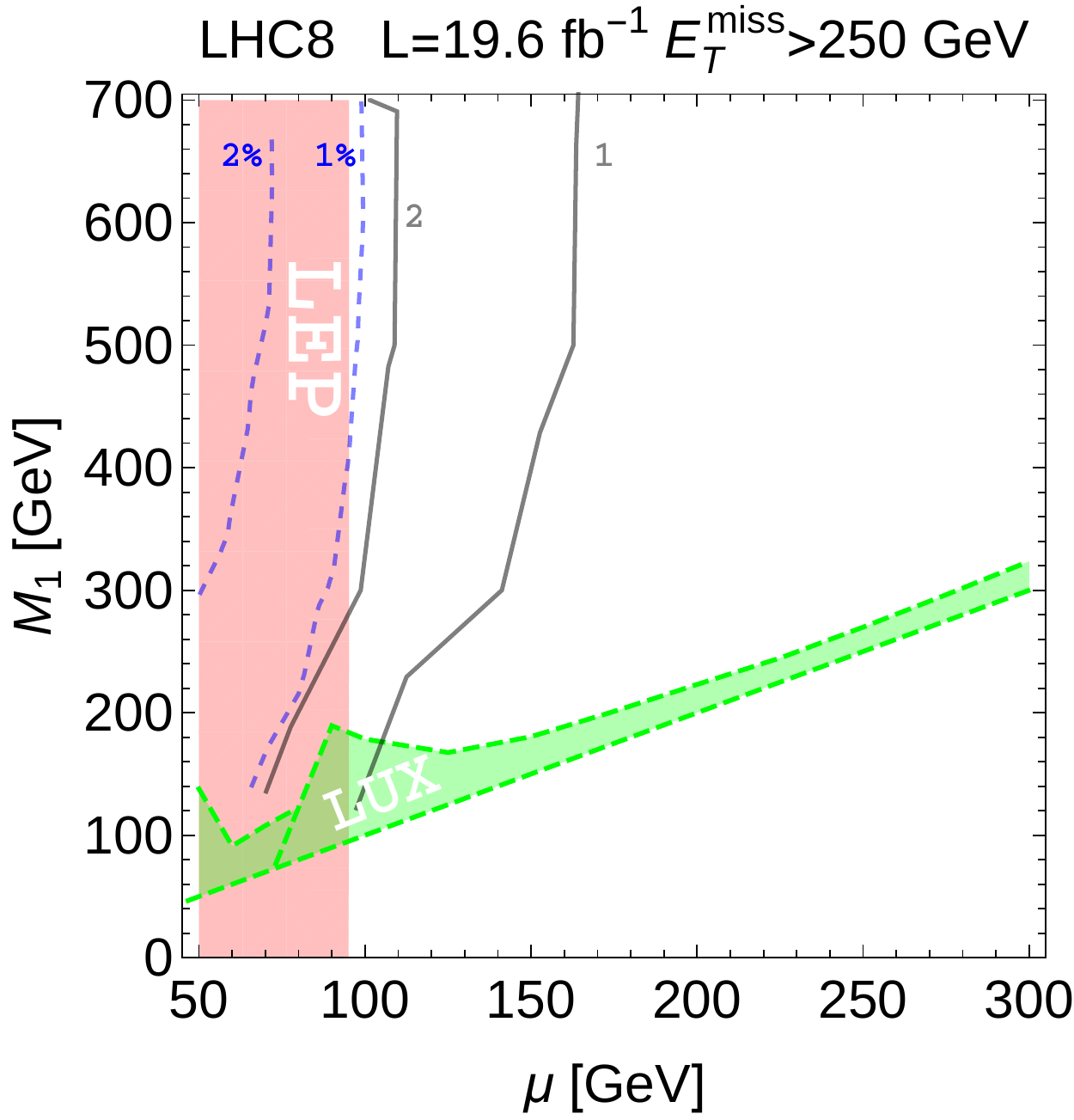}\hfill
\includegraphics[width=0.46\textwidth]{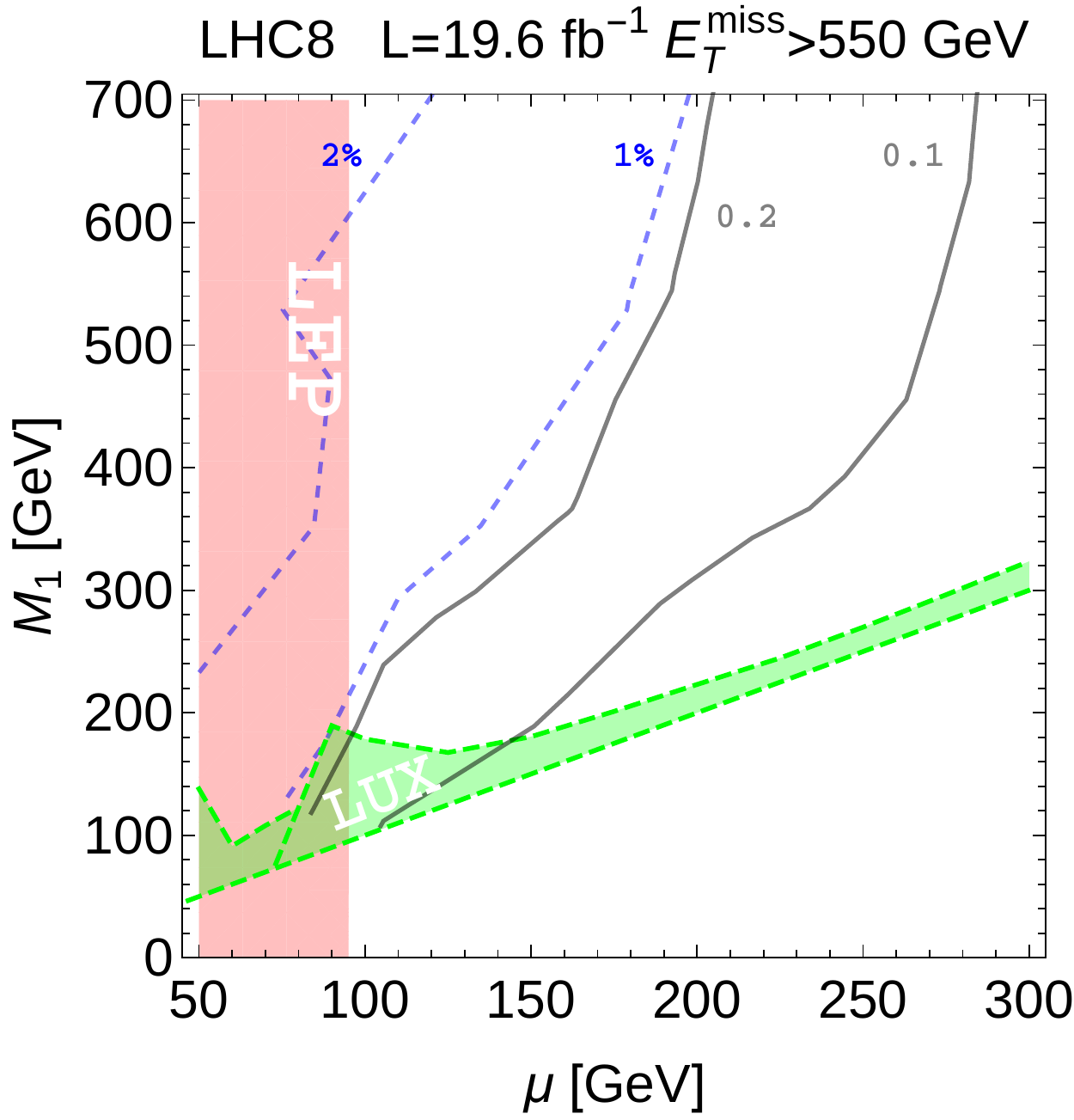}
\caption{Isocontours for $\alpha=2(\sqrt{S+B}-\sqrt{B})$ (gray) and S/B (blue-dashed) in the plane ($\mu$,$M_1$) for the signal regions 
1 and 7 as defined in~\cite{CMS:rwa}. LUX and LEP exclusions are shown in green and red respectively.
 }
 \label{fig:8tev}
\end{figure}

We then show in Fig.~\ref{fig:8tev} the resulting isocontours of the statistical significance, $\alpha=2(\sqrt{S+B}-\sqrt{B})$, and of the signal over background ratio, S/B, projected in the $\mu$--$M_1$ plane. The plots make clear that the results from the 8 TeV stage of the LHC are not able to set a bound on this SUSY scenario, due to the lack of statistical significance and, in the small region where $\alpha>2$, to the low S/B ratio, smaller than the actual systematic uncertainties of the analysis, which are of the order of 5--10\%. Nevertheless, an important feature is already clear, {\it i.e.} that the increase of the $E_T^{\rm miss.}$ cut has the capacity to improve the S/B ratio, especially for $\mu\gg M_1$. This however causes  a decrease of $\alpha$, that can anyway be, at least partially, compensated at the 13 TeV LHC by an increase of the integrated luminosity.

\subsection{13 TeV projections} 

As shown in the previous section, a higher cut on the $E_T^{\rm miss.}$ has the capacity of improving the S/B ratio, especially for small $\Delta M$, and in order to test how this can improve the reach of a monojet like analysis we have proceed as follows.
We have generated signal and background samples for the 13 TeV LHC with the same tools described in the previous section and we have then applied the following selection, inspired by the 8 TeV CMS analysis
\begin{itemize}
\item[-] Leading jet with $p_T>$ 200 GeV and $|\eta|<2.4$
\item[-] Veto on events with more than two jets with $p_T>$ 30 GeV and $|\eta|<$4.5
\item[-] $\Delta\phi(j_1,j_2)<$2.5
\item[-] Veto on electrons and muons with $p_T>$ 10 GeV.
\item[-] Veto on taus with $p_T>$ 20 GeV and $|\eta|<2.3$.
\end{itemize}
Finally, we have defined our signal regions with an increasing cuts on $E_T^{\rm miss.}$, from 200 GeV to 1000 GeV.

Given the tension between the increasing S/B ratio and decreasing $\alpha$ (for given integrated luminosity), with the increase of the $E_T^{\rm miss.}$ cut, it is important to identify an \emph{optimal cut} which, in the  $m_{\neut{0}}$--$\Delta M$ plane, can be defined as follows.

\begin{figure}[htb]
\centering
\includegraphics[width=0.5\textwidth]{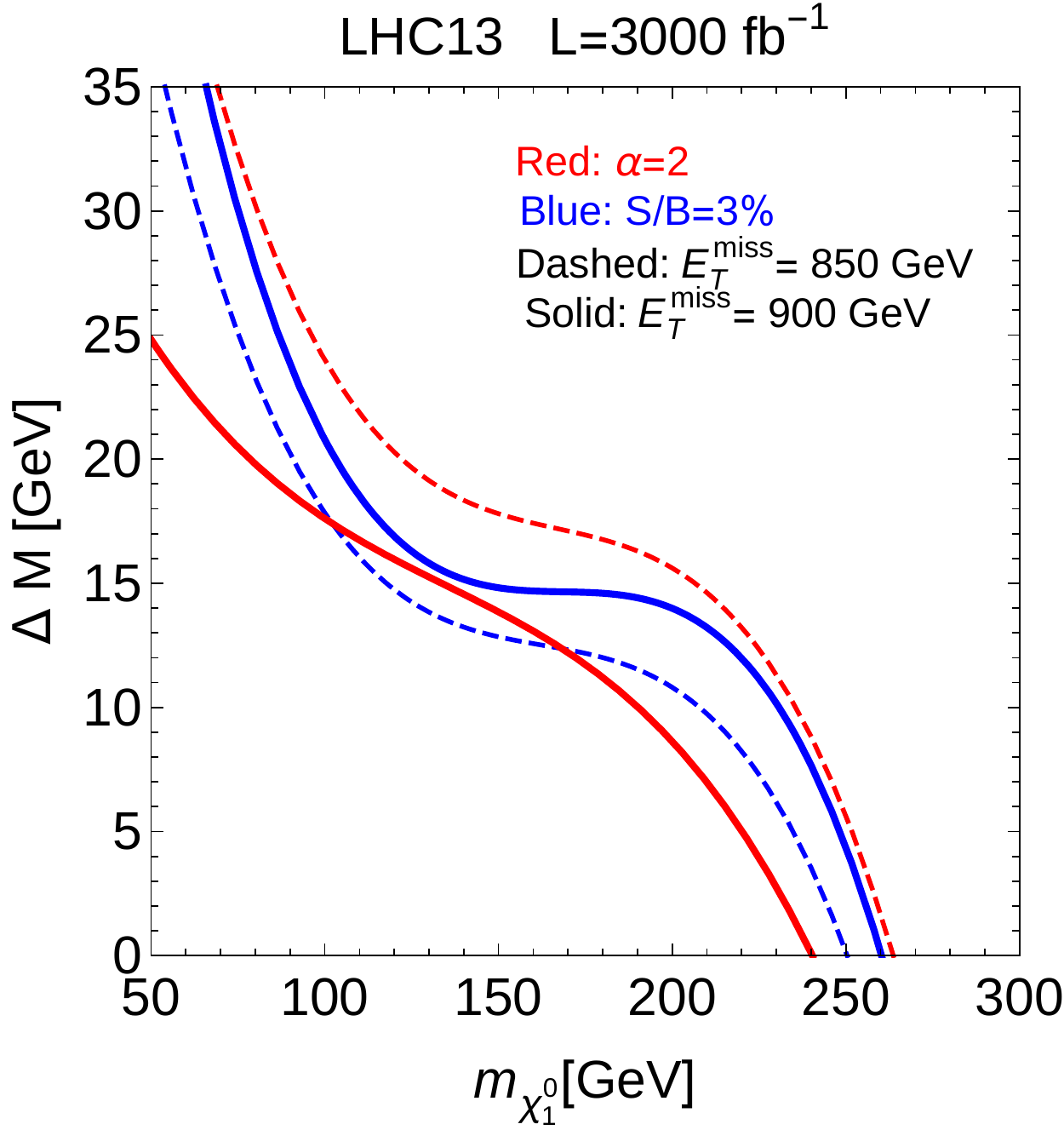}%
\caption{S/B (blue) and $\alpha$ (red) isocontours for two choices of $E_T^{\rm miss.}$ cut: 850 GeV (dashed) and 900 GeV (solid) in the $m_{\neut{1}}$-$\Delta $M plane.}
\label{sb-signif-tension}
\end{figure}
For a given value of integrated luminosity and S/B ratio, the optimal $E_T^{\rm miss.}$ cut   can be identified with the $E_T^{\rm miss.}$ value for which  the specific isocontours S/B (\emph{e.g.} S/B=3\%) and $\alpha$ (\emph{e.g.} $\alpha$=2 for exclusion or $\alpha$=5 for discovery) cross or are as close to each other as possible.
This is related to the fact that the isosignificance contours are shifted to the {\it left} in the $m_{\neut{1}}$-$\Delta$M plane with the increase in the $E_T^{\rm miss.}$ cut due to the {\it decrease} of signal statistics, while iso~S/B contours are shifted to the {\it right} at the same time. 
Therefore the case when the respective isocontours cross or are close to each other, would  provide the {\it maximal}  exclusion or discovery area in  the $m_{\neut{1}}$-$\Delta M$ plane, for a given integrated luminosity and minimum requirement on S/B.

We illustrate this in Fig.~\ref{sb-signif-tension},
which presents S/B and significance isocontours in the $m_{\neut{1}}$-$\Delta M$
plane for two different cuts on $E_T^{\rm miss.}$, 850 and 900 GeV, for 3000 fb$^{-1}$ of integrated luminosity. One can see that indeed for $E_T^{\rm miss.}>850$~GeV, the exclusion area
is below  $S/B=3\%$ (blue dashed) contour, 
while for  $E_T^{\rm miss.}>900$~GeV, the area below  $\alpha=2$ (red solid) contour
is excluded. Since for the first case the exclusion area is bigger,
the $E_T^{\rm miss.}>850$~GeV requirement is a better choice for the optimal cut.
 We have found that a cut around 600 (850) GeV for 100 fb$^{-1}$ (3000 fb$^{-1})$  provides
 $\alpha\simeq 2$ and $S/B\simeq 0.03$ isocontours optimally close to each other, which maximises the reach of the 13 TeV LHC. The proximity of $\alpha\simeq 2$ and $S/B\simeq 0.05$ isocontours requires instead a higher $E_T^{\rm miss.}$ cut which is found to be around 950 GeV, 
 and as a result leads to a poorer 13 TeV LHC reach.

\begin{figure}[htb]
\includegraphics[width=0.46\textwidth]{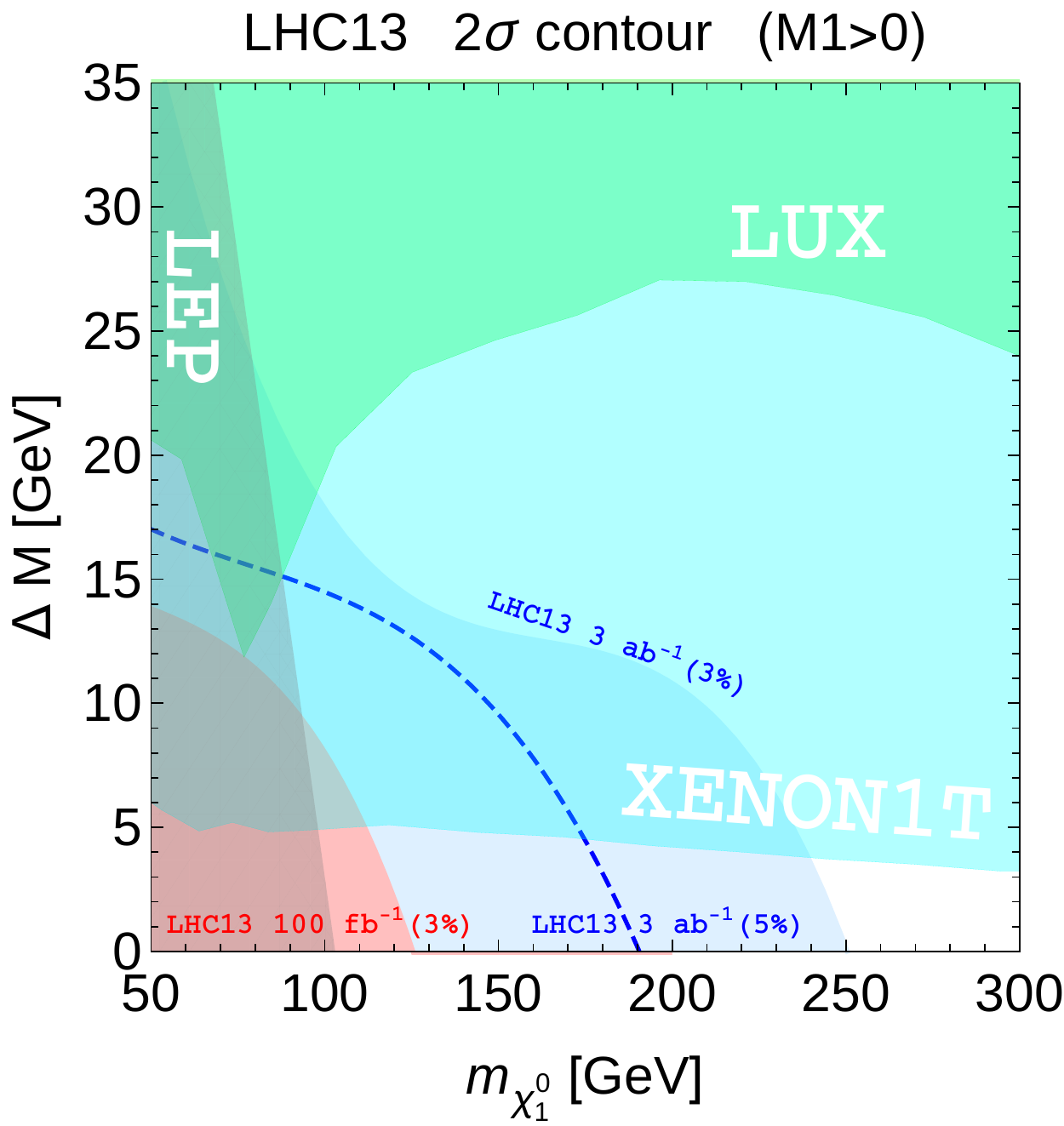}\hfill
\includegraphics[width=0.46\textwidth]{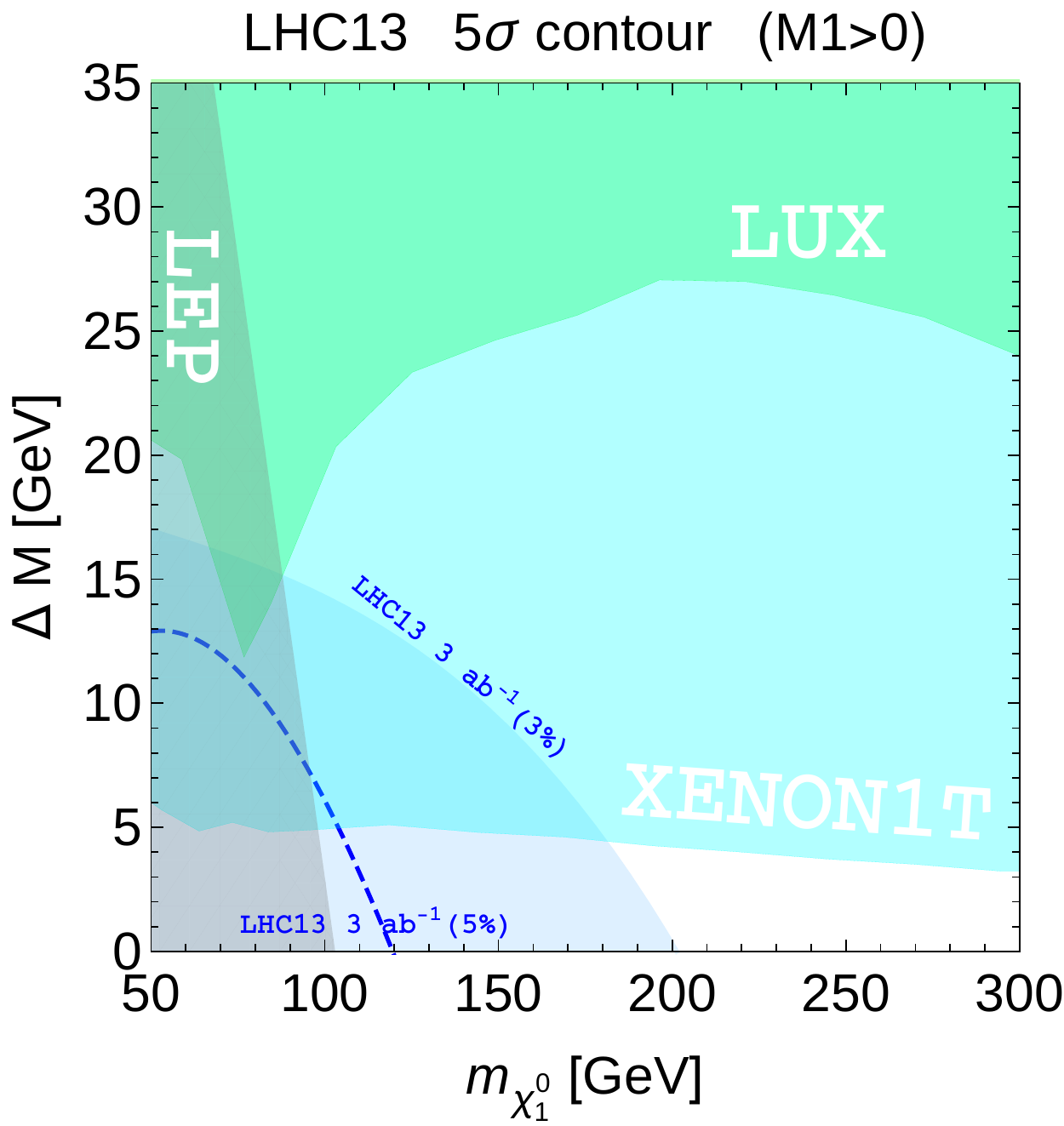}
\caption{Exclusion (left) and discovery (right) contour lines for the 13 TeV LHC at the end with 100 fb$^{-1}$ (light red region) and 3000 fb$^{-1}$ (light blue region) of integrated luminosity assuming S/B$>$3\%. For the latter case also the case S/B$>$5\% is shown. The region excluded by LUX and the projected exclusion 
by XENON1T are also shown, together with the LEP limit on the $\charpm{1}$ mass. $M_1>\mu$ is considered here.}
\label{fig:13tev}
\end{figure}

After the \emph{optimal cut} has been fixed for a given integrated luminosity and S/B ratio that one wants to achieve,  the reach of the 13 TeV run of the LHC can be easily presented in the $m_{\neut{0}}$--$\Delta M$ plane, and we illustrate this in Fig.~\ref{fig:13tev} for the case $M_1>0$.
We show both the cases where we require a value of 3\% and 5\% on the S/B ratio, the latter just for a high luminosity LHC option, while the former also for an integrated luminosity of 100 fb$^{-1}$. Also presented in the plot are the current reach of the LUX experiment and the projection for XENON1T, which clearly point out the complementarity between DD experiments, sensitive to high values of $\Delta M$, and the LHC, which will be able to test via monojet analyses a region with a small mass splitting.
In particular, while with 100 fb$^{-1}$ of integrated luminosity the 2$\sigma$ reach on $m_{\neut{1}}$ is $\sim$ 120 GeV assuming S/B$>$3\%, at the end of the high luminosity LHC program up to 250 GeV LSP can be tested. This value goes down to 200 GeV in case that a higher S/B ratio, $>$ 5\%, is required. Finally, with 3 ab$^{-1}$ of collected luminosity, the same scenario can be probed at a 5$\sigma$ level up to $\sim$ 200 GeV $\neut{1}$.

\section{Conclusion}

In this work we have explored the interplay of the LHC and DD underground experiments to probe DM signals in a natural SUSY scenario. In particular, after presenting the reach of the first run of the LHC, we have shown the projection for the 13 TeV stage of the CERN machine for two integrated luminosity options: 100 and 3000 fb$^{-1}$, taking into account realistic estimation on the determination of the experimental systematic uncertainties. 
Taking into account also the prospect for future DD experiments, we have highlighted the complementarity between the two approaches, showing that collider searches will be able to probe, via monojet like analyses, the region with small $\Delta M$, while underground experiments that with larger mass splitting.

\end{document}